\documentclass[conference,10pt]{IEEEtran}
\IEEEoverridecommandlockouts

\usepackage[utf8]{inputenc} %

\usepackage[T1]{fontenc} %

\usepackage{amsmath}
\usepackage{graphicx} %

\usepackage[
  backend=biber,
  style=ieee,
  sorting=none,
  url=false,      %
  doi=false,      %
  isbn=false,     %
  eprint=false,   %
  giveninits=true
]{biblatex}
\AtEveryBibitem{%
  \clearfield{url}%
  \clearfield{urlyear}%
  \clearfield{urlmonth}%
  \clearfield{urldate}%
  \clearfield{eprint}%
  \clearfield{note}%
  \clearfield{issn}%
  \clearfield{isbn}%
}

\usepackage[braket, qm]{qcircuit} %

\usepackage{chemformula} %
\setchemformula{
    format = \rmfamily,
}

\usepackage[hidelinks]{hyperref} %
\usepackage{cleveref}
\crefformat{subsection}{Subsection~#2#1#3}
\Crefformat{subsection}{Subsection~#2#1#3}

\usepackage{orcidlink} %
\usepackage{tabto} %

\usepackage{siunitx} %
\usepackage{bbold} %
\usepackage{mathtools} %
\usepackage{amssymb}
\usepackage{amsthm}
\usepackage{tikz}
\usepackage{cuted} %
\usepackage{bbm} %
\usepackage{enumitem} %

\usepackage{subcaption}
\usepackage{caption}
\usepackage{booktabs}
\usepackage{stfloats}
\usepackage{placeins}
\usepackage{float}
\usepackage{multirow} %
\usepackage[shortcuts,acronym]{glossaries} 
\setkeys{glslink}{hyper=false} %

\newacronym{CB}{CB}{Conduction Band}
\newacronym{CW}{CW}{Continuous Wave}
\newacronym{FSS}{FSS}{Fine Structure Splitting}
\newacronym{MAC}{MAC}{Message Authentication Code}
\newacronym{QD}{QD}{Quantum Dot}
\newacronym{QC}{QC}{Quantum Channel}
\newacronym{QKD}{QKD}{Quantum Key Distribution}
\newacronym{TPE}{TPE}{Two-Photon Excitation}
\newacronym{VB}{VB}{Valence Band}
\newacronym{CPTP}{CPTP}{completely positive, trace-preserving}
\newacronym{CCR}{CCR}{canonical commutation relation}
\newacronym{POVM}{POVM}{positive operator-valued measure}
\newacronym{HOM}{HOM}{Hong-Ou-Mandel}
\newacronym{QIO}{QIO}{quantum input-output}

\DeclareMathOperator{\Tr}{Tr} %

\newcommand{\ie}{\textrm{i.e.}~} %

\newcommand{\eg}{\textrm{e.g.}~} %

\DefineBibliographyStrings{english}{%
    andothers = {\textrm{et al\adddot}}
}

\bibliography{bibliography}

\author{
    \IEEEauthorblockN{Simon Sekavčnik\IEEEauthorrefmark{1}\IEEEauthorrefmark{2},
    Janis Nötzel\IEEEauthorrefmark{2}}
    \IEEEauthorblockA{\IEEEauthorrefmark{1}Corresponding author: simon.sekavcnik@tum.de}
    \IEEEauthorblockA{\IEEEauthorrefmark{2}Emmy-Noether Group Theoretical Quantum Systems Design,\\
    TUM School of Computation, Information and Technology,\\
    Technical University of Munich (TUM), Germany}
    \thanks{Copyright © 2026 IEEE. This material may be used for personal use only. Any other use requires permission from IEEE. This work has been accepted for publication at the IEEE International Conference on Quantum Communications, Networking, and Computing (QCNC 2026).} 
}

\begin{document}
\title{Symbolic Quantum State Representation and its Simulation}
\maketitle

\vspace{0.5em}

\begin{abstract}
We introduce a symbolic operator framework for simulating quantum photonic systems that works directly with the canonical commutation relations and the Weyl algebra. Unlike existing Fock-space or Gaussian simulators, our method treats temporal wave packets and polarization modes in a continuous setting and does not rely on discretization or Hilbert-space truncation. Device operations are expressed as algebraic rewrite rules acting on creation and annihilation operators, allowing exact evolution of finite-photon states through linear optical networks. As an illustration, we reproduce Hong-Ou-Mandel interference for Gaussian pulses with controlled temporal and spectral mismatch.
\end{abstract}

\section{Introduction}
The ability to accurately simulate quantum photonic systems is central to both fundamental science and emerging quantum optical technologies. Photonic platforms are widely explored for quantum communication \cite{strobelTelecomwavelengthQuantumTeleportation2025}, metrology \cite{barbieriOpticalQuantumMetrology2022}, and computing \cite{kokReviewArticleLinear2007}. 

Existing simulation frameworks typically adopt one of two paradigms. \emph{Fock-space} based approaches (\eg QuTiP\cite{lambertQuTiP5Quantum2024a}, Strawberry Fields\cite{killoranStrawberryFieldsSoftware2019}, PhotonWeave\cite{sekavcnikPhotonWeave2025}) provide a direct matrix representation of states and operators, enabling general-purpose simulations including non-Gaussian states.
However, these methods rely on a orthogonal, truncated Fock basis. 
On the other hand, \emph{Gaussian simulators} (covariance-matrix based methods, continuous-variable libraries) exploit the efficient description of Gaussian states under linear optics and squeezing (The Walrus \cite{guptWalrusLibraryCalculation2019}). These tools enable large-scale simulations but break down when non-Gaussian resources, spectral distinguishability, or mode mismatch are introduced.

Alternative treatments of quantum optical systems are provided by \gls{QIO} theory \cite{gardinerInputOutputDamped1985}, which models optical channels using continuous-time field operators. It provides elegant analytical tools for describing open-system dynamic and scattering processes. \gls{QIO} theory, however, is formulated as a theoretical framework rather than a general-purpose simulator. It does not track arbitrary wave-packet shapes or produce explicit multi-photon states for interferometric networks.

Even though all the aforementioned simulation approaches enable rigorous treatment of the optical fields, there is no simulation framework that unifies the non-Gaussian state evolution with a continuous-mode description of optical fields. Existing Fock-space simulators require discretization by enforcing orthogonality. Gaussian simulators can represent arbitrary wave-packet shapes, but only after expressing them in an orthonormal discrete basis. They therefore cannot operate directly on non-orthogonal continuous-mode wave-packet labels, and they remain restricted to states with Gaussian statistics. \gls{QIO} theory offers continuous-time field operators but does not produce explicit multi-photon output states for interferometric transformations.

In this work, we introduce a symbolic operator framework that fills this gap by acting directly on the \gls{CCR} and underlying Weyl algebra of photonic modes \cite{bratteliOperatorAlgebrasQuantum1997}. Optical modes are represented in a global Fock space constructed from symmetrized single-photon spaces whose labels include both temporal wave-packet and polarization degrees of freedom. Device actions are implemented as algebraic rewrite rules (defined in Sec.~\ref{sec:device_maps}) acting on ladder operators. This approach enables exact, discretization-free propagation of arbitrary finite-photon quantum optical states. We propose a general quantum simulation prototype and demonstrate the approach on the \gls{HOM} interference for Gaussian wave packets with temporal and spectral mismatch.

\section{Mathematical framework}\label{sec:mathematical_framework}
This section describes the Hilbert space which underpins the representation of optical states under study.

\subsection{Single photon space}
We start by describing the state of a single photon in a single optical channel (\eg fiber). The state space is characterized as a vector in a Hilbert space:
\begin{equation}
    \mathcal{H}_1 = \mathcal{H}_\mathrm{env}\otimes\mathcal{H}_\mathrm{pol},
\end{equation}
which combines a temporal degree of freedom with a polarization degree of freedom. Through Fourier transform this equivalently represents the spectral profile and via propagation at the speed of light also determines spatial dependence along the optical path \cite{zhangControlContinuousmodeSinglephoton2021}.

In such description, the choice for the $\mathcal{H}_\mathrm{env}$ space is the space of square-integrable complex functions $L^2(\mathbb{R}, \mathbb{C})$ \cite{zhangControlContinuousmodeSinglephoton2021, brechtPhotonTemporalModes2015, raymerTemporalModesQuantum2020}. The space already includes an inner product definition:
\begin{equation}
    \langle f, g \rangle_\text{env} = \int_{\mathbb{R}} f^*(t)g(t)\, dt.
\end{equation}

The polarization Hilbert space $\mathcal{H}_\text{pol}$ models the internal spin degree of freedom for a single photon. In the paraxial approximation the polarization behaves as a two-level quantum system described as a Hilbert space $\mathcal{H}_\text{pol}=\mathbb{C}^2$ as implied in Eq.~(9.22) in \cite[p.~221]{gerryIntroductoryQuantumOptics2023}. A general polarization state can be written as:
\begin{equation}
    \ket{\phi_\text{pol}} = p_H\ket{H} + p_V\ket{V},
\end{equation}
with amplitudes $p_H, p_V \in \mathbb{C}$. The inner product is then defined as:
\begin{equation}
    \langle \phi_{\text{pol},1} | \phi_{\text{pol},2} \rangle_\text{pol} = p_{1,H}^*p_{2,H} + p_{1,V}^*p_{2,V}.
\end{equation}

The inner product of the single photon space can then be written as a multiplication of its constituents:
\begin{equation}
    \langle \psi_{1} \mid \psi_{2} \rangle_{\mathcal{H}_1} = \langle f_1, f_2\rangle_\text{env}\langle \phi_1 | \phi_2\rangle_\text{pol},
\end{equation}
for all $\psi_i = \ket{f_i}_\text{env}\otimes\ket{\phi_i}_\text{pol}\in \mathcal{H}_1$.
\subsubsection{Gaussian mode family}
Although $L^2(\mathbb{R}, \mathbb{C})$ contains states with for arbitrary temporal envelope, we focus on the Gaussian function family $\mathcal{G} \subset L^2(\mathbb{R}, \mathbb{C})$ as it provides a convenient closed-form analytic expression for overlaps and Fourier transforms, and is widely used in single-photon modeling.

Each temporal mode function $\zeta(t) \in \mathcal{G}(\mathbb{R}, \mathbb{C})$ is defined as:
\begin{equation}
    \zeta_{\mathbf{T}}(t) = \left(\frac{1}{2\pi\sigma^2}\right)^{1/4} e^{-\frac{(t-\tau)^2}{4\sigma^2}} e^{-i\omega_0(t-\tau)},
\end{equation}
where $\mathbf{T}=(\sigma,\tau,\omega_0)$ is a temporal mode label, specified by:
time shift $\tau$, central frequency $\omega_0$, and the temporal width $\sigma$. The factor $\left(\frac{1}{2\pi\sigma^2}\right)^{1/4}$ ensures the normalization in $L^2$, and global phase is omitted, since physically only relative phases between pulses or between modes matter \cite[Sec 2.2.7]{nielsenQuantumComputationQuantum2010}.

The normalized Gaussian mode family choice provides a closed analytic form for overlap $\langle \zeta_{\mathbf{T}_1}, \zeta_{\mathbf{T}_2} \rangle$. For two such Gaussians with parameters $(\tau_k, \omega_k, \sigma_k)$, the closed-form inner product expression simplifies to:
\begin{align}\label{eq:gauss_closed_form}
    \begin{split}
        \langle \zeta_{\mathbf{T}_1}, \zeta_{\mathbf{T}_2} \rangle = 
        \sqrt{\frac{2\sigma_1\sigma_2}{S^2}}\;
        &\exp\Bigg[-\frac{\Delta\tau^2}{4S^2}-\frac{\sigma_1^2\sigma_2^2}{S^2}(\Delta\omega)^2\Bigg]\\
        \times &\exp\Bigg[i\frac{\Delta\tau(\sigma_1^2\omega_1+\sigma_2^2\omega_2)}{S^2}\Bigg],
    \end{split}
\end{align}
with substitutions $S^2=\sigma_1^2+\sigma_2^2$, $\Delta\tau=\tau_1-\tau_2$, and $\Delta\omega=\omega_1-\omega_2$.

In the limiting cases $\sigma\to0$ and $\sigma\to\infty$, the Gaussian envelope tends to a delta pulse $\delta_t$ in time (flat spectrum) or to a monochromatic standing wave (spectral delta $\delta_\omega$). Neither of these limiting distributions belongs to $L^2$. They can still be approximated numerically, choosing very small or very large but finite values $0<\sigma<\infty$.

The Gaussian family $\mathcal{G}$ is chosen for its convenience and relevance, the mathematical formalism however, applies to a general mode function in $L^2(\mathbb{R},\mathbb{C})$.

\subsection{Fock space}

Photons in the same optical channel can be described with the use of Fock space. To model multiple optical channels (\ie distinct fibers), we label each such channel with $p\in \mathcal{P}$. To each such channel we associated separate single-photon space $\mathcal{H}_1^{(p)}$. Individual spaces associated with different optical channels are taken to be mutually orthogonal:
\begin{equation}
    \langle \psi^{(p)}, \phi^{(q)} \rangle = 0
    \qquad \text{for all } \psi^{(p)}\in\mathcal{H}_1^{(p)},\;
    \phi^{(q)}\in\mathcal{H}_1^{(q)},\; p\neq q.
\end{equation}
This reflects the physical assumption that photons propagating in distinct optical paths correspond to independent degrees of freedom.

For one channel labeled with $p$, we construct a Fock space \cite[p.7]{bratteliOperatorAlgebrasQuantum1997}:
\begin{equation}
\mathcal{F}^{(p)} = \bigoplus_{n=0}^\infty \text{Sym}^n(\mathcal{H}_1^{(p)}),
\end{equation}

where $\mathrm{Sym}^n$ denotes the $n$-fold symmetrized tensor product.

The global Hilbert state consisting of all optical channels is the tensor product of individual channel Fock spaces:
\begin{equation}\label{eq:global_fock_space}
    \mathcal{F}_\text{g} = \bigotimes_{p\in\mathcal{P}}\mathcal{F}^{(p)}.
\end{equation}

\section{Operator Algebra}\label{sec:operator_algebra}

Having defined the Hilbert space in which we operate, we continue with generating the canonical creation and annihilation operators and defining the relationship between them in the context of operators and states.

\subsection{Creation and Annihilation operators}
We define creation  and annihilation operators associated with arbitrary temporal, polarization and path space. These operators generate algebra of optical observables used throughout the simulation framework.

Each single-photon mode in a given path $p$ is fully described by a vector $\psi^{p} \in \mathcal{H}^{(p)}_1$. The continuous-mode operators obey:
\begin{equation}
    [\hat{a}_{\phi, p}(t), \hat {a}^\dagger_{\phi',p'}(t')] = \delta_{pp'} \delta_{\phi\phi'}\delta(t-t').
\end{equation}

These operators create and destroy photons with profile $f$, polarization $\phi$ in path $p$. For brevity we combine the path label with mode description as:

\begin{equation}
\mathcal{M} = \{\, (p,f) \mid p\in\mathcal{P},\; f \in \mathcal{H}^{(p)}_1 \,\}.
\end{equation}

where each element is written as $m=(p,f)$. A mode label $m\in \mathcal{M}$ therefore denotes both single-photon profile $f$ and the spatial channel $p$. The smeared ladder operator pair can then be rewritten simply as $\hat{a}(m)$ and $\hat{a}^\dagger(m)$:
\begin{align}
    \hat{a}(m_i) &= \int f_\text{env,i}^*(t)\hat{a}_{\phi_i, p_i}(t) dt,\\
    \hat{a}^\dagger(m_i) &= \int f_\text{env,i}(t)\hat{a}^\dagger_{\phi_i, p_i}(t) dt,
\end{align}
Here $f_\text{env}\in L^2(\mathbb{R}, \mathbb{C})$ is a component of $\mathcal{H}_1$.

Following Bratteli \& Robinson \cite[p.10]{bratteliOperatorAlgebrasQuantum1997}, we then also define the \gls{CCR} between individual ladder operators:
\begin{align} \label{eq:ccr}
    \begin{split}
    [\hat{a}(m_i), \hat{a}(m_j)]=[\hat{a}^\dagger(m_i), \hat{a}^\dagger(m_j)]=0,\\
    [\hat{a}(m_i), \hat{a}^\dagger(m_j)]= \langle m_i, m_j \rangle \mathbbm{1},
    \end{split}
\end{align}
where the inner product follows from the definitions in the previous sections:
\begin{equation}
    \langle m_i, m_j \rangle = \delta_{p_i, p_j} \langle \psi_i, \psi_j\rangle_{\mathcal{H}_1},
\end{equation}
so that the modes in different spatial channels are orthogonal, while the overlaps within one such channel are dictated by the overlaps of the corresponding single-photon profile.

These ladder operators together with their commutation relations generate the standard CCR algebra for bosonic modes \cite[Sec.~5.2.2]{bratteliOperatorAlgebrasQuantum1997}. All observables in our framework are built from polynomial combinations of these operators.

\subsection{Operator Algebraic Structures}

The \gls{CCR}, we introduced above in Eq.~\eqref{eq:ccr}, fully determines how individual ladder operators interact. To describe the general relevant transformations, we must work with expressions built from multiple ladder operators that form a polynomial.

Utilizing the construction of \cite{bratteliOperatorAlgebrasQuantum1997}, we introduce basic operator expressions, the notion of normal ordering and the polynomial operator structure, which we use throughout the simulation framework. 

\subsubsection{Operator Monomials}

A monomial is a finite product of ladder operators with arbitrary ordering. Given mode labels $m_1, \ldots m_k \in \mathcal{M}$, a general monomial is formed as:
\begin{equation}
    A = \hat{a}(m_{i_1}) \ldots \hat{a}(m_{i_r})\hat{a}^\dagger(m_{j_1}) \ldots \hat{a}^\dagger(m_{j_s}),
\end{equation}
where individual ladder operators appear in any order. An empty product is interpreted as the identity operator.

\subsubsection{Normal Ordering}

Normal ordering is a rewriting procedure on a monomial, where the monomial is rewritten by repeatedly applying the \gls{CCR} rule, which moves all creation operators to the left of the annihilation operators:

\begin{align}\label{eq:normal_ordering}
   [\hat{a}(m_i), \hat{a}^\dagger(m_j)]= \hat{a}(m_i)\hat{a}^\dagger(m_j) - \hat{a}^\dagger(m_j)\hat{a}(m_i)\\
   \hat{a}(m_j)\hat{a}^\dagger(m_i)=\hat{a}^\dagger(m_i)\hat{a}(m_j) + \langle m_i, m_j \rangle\mathbb{1}.
\end{align}

We denote an already normally ordered monomial as $M(I|J)$. That is, a monomial which has all creation operators on the left of all annihilation operators. Where $I$ is represented by a list of creation operators and $J$ is the list of annihilation operators and the notation $\emptyset$ represents the empty list. Further, $M(\emptyset|\emptyset) = \mathbbm{1}$ is taken to be an identity operator.

\subsubsection{Polynomial Operators}

We define a polynomial operator as a linear combination of normally ordered monomials:
\begin{equation}
    P = \sum_{(I,J)} c_{I,J}M(I|J), \quad c_{I,J} \in \mathbb{C}.
\end{equation}

Then the adjoint of such polynomial operator is:
\begin{equation}
    P^\dagger = \sum_{(I,J)} c^*_{I,J}M(J|I), \quad c_{I,J} \in \mathbb{C}.
\end{equation}

In order to make normal form unique, we assume some arbitrary but fixed order based on the mode label $m\in\mathcal{M}$. This ordering rule enforces the order of the ladder operators in monomials.

\subsection{Quantum States}
Only once we have the polynomial operators defined we can fully express the quantum states in the global Fock space from Eq.~\ref{eq:global_fock_space}. States by definition consist only of creation operators. As such any ket state is represented as a polynomial:
\begin{equation}\label{eq:pure_state}
    \ket{\psi} = \sum_I c_I M(I|\emptyset) \ket{0}, \quad \ket{\psi} \in \mathcal{F}_g.
\end{equation}

In the same way we also define a density operator for the state as:
\begin{equation}\label{eq:density_matrix}
    \rho = \sum_{I,J} c_{I,J} M(I|\emptyset)\ket{0}\bra{0}M(J|\emptyset)^\dagger.
\end{equation}

Note that the left and right monomials $M(I|\emptyset)$ and $M(J|\emptyset)$ are, in general, different normally ordered monomials, but they both contain creation operator only prior to adjoint.

\subsection{Contraction rules}

To define different operations on states and operators we clarify the basic ruleset. Firstly, it is important to mention that applying an annihilation operator to a vacuum state destroys the state:
\begin{equation}
    \hat{a}(m_i)\ket{0} =0, \qquad \bra{0}\hat{a}^\dagger(m_i) = 0.
\end{equation}

To compute the inner product we insert the monomial expansions of the states Eq.~\eqref{eq:pure_state} into $\ip{\phi}{\psi}$:
\begin{equation}\label{eq:inner_product}
    \langle \phi|\psi \rangle = \sum_{I,I'} c^*_I d_{I'} \langle0|\hat{a}(m_{I_1})\ldots \hat{a}(m_{I_r})
    \hat{a}^\dagger(n_{I'_1})\ldots\hat{a}^\dagger(n_{I'_s})|0\rangle.
\end{equation}

For each pair of monomials this produces an operator string whose vacuum expectation value is evaluated by normal ordering. Each commutation of an annihilation past a creation operator yields a contraction $\langle m_i, m_j \rangle\mathbb{1}$, and only fully contracted terms survive (see Appendix.~\ref{appendix:contraction_examples}).

\subsection{Trace and Partial Trace}

The canonical trace operation on a density operator can be expressed as a weighted sum of vacuum
inner products of normally ordered monomials:
\begin{equation}
    \Tr(\rho)=\sum_{I,J} c_{I,J}\langle 0 | M(J|\emptyset)^\dagger M(I|\emptyset) | 0\rangle.
\end{equation}

To evaluate this expression, again, the product $M(J|\emptyset)^\dagger M(I|\emptyset)$ has to be contracted (normally ordered) and only the fully contracted terms contribute to the final trace value. 

Partial trace is similar to regular trace, the difference is that trace is done over some subspace $\mathcal{F}_E$. The idea is to divide the full Fock space into a system $S$ and environment $E$:
\begin{equation}
    \mathcal{F}_g = \mathcal{F}_S \otimes \mathcal{F}_E.
\end{equation}

This splitting in the space is then represented by rewriting each monomial for a state in a product of system and environment operators:
\begin{equation}
M(I|\emptyset) = M^{(S)}(I_S|\emptyset)M^{(E)}(I_E|\emptyset),
\end{equation}
and similarly for $M(J|\emptyset)$.

We then obtain the reduced state by evaluating the vacuum expectation of all environment operators, as in the case of full trace:
\begin{equation}\label{eq:partial_trace}
\Tr_E(\rho) = \sum_{I,J} c_{I,J}M^{(S)}(I_S|\emptyset)\bra{0_E} P_E\ket{0_E} M^{(S)}(J_S|\emptyset)^\dagger,
\end{equation}
where:
\begin{equation}
    P_E=M^{(E)}(J_E|\emptyset)^\dagger M^{(E)}(I_E|\emptyset)
\end{equation}
denotes the environment operator product whose vacuum expectation value yields the contraction factor. The process of evaluating inner product and vacuum contractions is illustrated in appendix \ref{appendix:contraction_examples}.

\subsection{Observables and Expectation Values}
Operator polynomials are defined more generally than state polynomials in Eq.~\eqref{eq:pure_state} and can also include the annihilation operators.

An observable represented as operator polynomial
can be expanded in the normally ordered monomial basis $M(I|J)$,
\begin{equation}
    O = \sum_{I,J} c_{I,J} M(I|J),
\end{equation}
and its expectation value in a density state $\rho$ is defined as
\begin{equation}
    \langle O \rangle_\rho=\text{Tr}(\rho O).
\end{equation}

This trace is evaluated by multiplying $\rho$ and $O$ terms followed by contraction. This automatically includes all degrees of freedom of the non-orthogonal basis modes.

\section{Device maps}\label{sec:device_maps}
Until now we have presented the framework, operators and rules that guide the manipulation of these operators. Now we shift our attention to some specific devices and outline their action on quantum states. In our model, we formulate device actions in the Heisenberg picture, where each component acts as a map that transforms creation and annihilation operators, while the vacuum state these operators act on remains fixed:

\begin{equation}\label{eq:algebra_map}
    \Phi : \mathcal{A} \to \mathcal{A}.
\end{equation}

Here $\mathcal{A}$ denotes the polynomial \gls{CCR} algebra generated by ladder operators and finite linear combinations of normally ordered monomials introduced in Sec.~\ref{sec:operator_algebra}.

In this section we outline basic set of devices commonly used in the laboratory setups.

\subsection{Sources}

Within this algebraic formalism, optical sources are represented as device maps, application of which leading to the generation of excitations by introducing creation operators.

Sources constitute a special type of transformation in that source maps do not define algebra homomorphisms on $\mathcal{A}$. Instead sources operate by pre-pending (and then normal ordering) creation operators onto an existing state, thereby generating additional excitations. This means that source maps act directly on the global Fock space:
\begin{equation}
    \Sigma: \mathcal{F}_g \to \mathcal{F}_g.
\end{equation}

A deterministic \textbf{single-photon source} prepares one photon state in a defined mode,
\begin{equation}
    \Sigma_\text{SPS}(\lvert \psi \rangle) = \hat{a}(m_i)^\dagger \lvert \psi \rangle,
\end{equation}
followed by normal ordering. Here the mode label $m_i$ describes the mode of the produced excitation.

This recipe can be used to model arbitrary finite-photon number type of source. A \textbf{photon pair sources}, for example,  such as spontaneous parametric down-conversion or biexciton-exciton cascade \cite{sekavcnikEntangledPhotonPair2025} generate correlated photons,
\begin{equation}
    \Sigma_\text{pair}(\lvert \psi \rangle)=\sum_{ij}f_{ij}\hat{a}(m_i)^\dagger \hat{a}(m_j)^\dagger\lvert \psi \rangle,
\end{equation}
where coefficients $f_{ij}$ represent the joint spectral or polarization amplitude of the emission.

Because the algebra is closed under creation and annihilation operators, all these states can be expressed as polynomials acting on the vacuum.

\subsection{Linear (passive) devices}\label{sec:linear_passive_devices}
Passive linear devices, such as beam splitters, phase shifters, and polarization rotators, correspond to \textbf{unitary transformations} acting on an ordered mode tuple $(a_1, \ldots, a_n)$. These maps are maps on the operator algebra outlined in Eq.~\eqref{eq:algebra_map}.

A transformation matrix $U\in\mathbb{C}^{n\times n}$ defines the action:
\begin{align}
    \begin{split}
    \Phi(\hat{a}^\dagger(m_k)) &=\sum_j U^*_{jk}\hat{a}^\dagger(m_j),\\
    \Phi(\hat{a}(m_k)) &=\sum_j U_{jk}\hat{a}(m_j),
    \end{split}
\end{align}
which preserves the canonical commutation relations since
\begin{equation}
    \left[\Phi\left(\hat{a}\left(m_i\right)\right), \Phi\left(\hat{a}^\dagger\left(m_j\right)\right)\right] = \sum_{k,l}U_{ki}U^*_{lj}\langle m_k, m_l\rangle.
\end{equation}

This type of transformation preserves the \gls{CCR} if it leaves the inner products between modes invariant.
\begin{equation}
    \sum_{k,l}U_{ki}U^*_{lj}\langle m_k, m_l\rangle = \langle m_i, m_j\rangle.
\end{equation}
In the special case of orthonormal modes this reduces to the usual unitarity condition $UU^\dagger=\mathbbm{1}$.

The typical examples of the passive linear devices include \cite[p.187-191]{leonhardtQuantumStatisticsLossless1993}:
\begin{itemize}
    \item \textbf{Phase shifter} acting on one mode:
    \begin{equation}
        \hat{a}^\dagger(m'_i) = e^{i\phi}\hat{a}^\dagger(m_i).
    \end{equation}
    \item \textbf{Beam splitter} acting on two modes :
    \begin{equation}
        \begin{pmatrix}
            \hat{a}(m_1')\\\hat{a}(m'_2)
        \end{pmatrix} = 
        \begin{pmatrix}
            c & s \\ -s & c
        \end{pmatrix}
        \begin{pmatrix}
            \hat{a}(m_1)\\\hat{a}(m_2)
        \end{pmatrix}
        , \quad \lvert c \rvert^2+\lvert s \rvert^2 = 1.
    \end{equation}
    \item \textbf{Polarization rotator} acting on polarizations:
    \begin{equation}
        \begin{pmatrix}
            a_H'\\a_V'
        \end{pmatrix}=R(\theta) \begin{pmatrix}
            a_H \\a_V
        \end{pmatrix}, \quad R(\theta) = \begin{pmatrix}
            \cos\theta & \sin\theta\\
            -\sin\theta & \cos\theta
        \end{pmatrix}
    \end{equation}
\end{itemize}

\subsection{Lossy, filtering, and amplifying channels}
Lastly, we introduce a class of linear devices which mix the system modes with the environmental vacuum modes. Any such transformation can be written as a Bogoliubov transformation acting on the extended set of modes:
\begin{equation}
    \Phi_{b}\left( \hat{a}(m_i)\right) = \sum_j U_{ij}\hat{a}(m_j) + \sum_j V_{ij}\hat{a}^\dagger(m_j),
\end{equation}
with additional requirement, that the \gls{CCR} be preserved:

\begin{equation}
    \sum_{k,l}\left(U_{ik}U^*_{jl}-V_{ik}V^*_{jl}\right)\langle m_k,m_l\rangle = \langle m_i,m_j\rangle.
\end{equation}

\subsection{Lossy channels}
A frequency independent lossy channel with transmitivity $0\leq\eta\leq1$ acts as:
\begin{equation}
    \Phi_\text{loss}(\hat{a}(m_i))= \sqrt{\eta}\hat{a}(m_i)+\sqrt{1-\eta}\hat{e}(m_i).
\end{equation}

In the Bogoliubov representation introduced above, this corresponds to the case $V=0$, \ie no mixing with environmental creation operators.

The auxiliary mode $\hat{e}(m_j)$ represents the vacuum fluctuations entering through the open port. When tracing out the environment, the system state exhibits exponential decay of mean photon number and reduced coherence, reproducing the standard form of optical damping found in master equation approaches.

This model is also the algebraic analogue of the beam-splitter model of loss \cite[Sec 7.5]{loudon2000quantum}, where the transmitted and lost components correspond to the two output ports of an imaginary beam splitter.

\subsubsection{Filtering channels}
Spectral and polarization filters are modeled as \textbf{non-unitary maps} that attenuate or remove part of the optical field. A \gls{CPTP} map describing the filtering process is described by:

\begin{equation}
    \Phi_\text{F}\left(\hat{a}(m_i)\right) = T_i \hat{a}(m'_i) + \sqrt{1-\lvert T_i\rvert^2}\;\hat{e}(m_i),
\end{equation}
where the environmental operators again satisfy the relation $[\hat{e}(m_i),\hat{e}^\dagger(m_j)]=\delta_{ij}$. The mode label $m'_i$ labels the mode associated with the reshaped modal envelope $f'_i(\omega)$ or polarization $\ket{\phi_\text{pol}}$. 
In the Bogoliubov representation this again corresponds to $V=0$, since filtering attenuates the mode without introducing environmental excitations.

In the case of \emph{spectral filters}, such operation also transforms the pulse envelope according to the filter transfer function $T(\omega)$:

\begin{equation}
    f_i'(\omega) = T(\omega)f_i(\omega).
\end{equation}

The scalar coefficients $T_i$ correspond to the projection of the reshaped and appropriately normalized envelope $f_i'(\omega) \in L^2(\mathbb{R}, \mathbb{C})$ back onto the original mode envelope $f_i(\omega)$:
\begin{equation}
    T_i = \langle f_i , f'_i\rangle = \int T(\omega)|f_i(\omega)|^2 \; d\omega, 
\end{equation}
the remaining orthogonal component is placed into the corresponding environmental mode.

Another type of filter we analyze here is \emph{polarization filter}, which works in similar way as the spectral filter above, but as the name suggests, it acts on the polarization components. Here the transfer function is not a spectral function but a linear operator acting on the polarization subspace $\mathcal{H}_\text{pol}$. We denote the transfer map as $T_\text{pol}$, which must obey $T_\text{pol}^\dagger T_\text{pol} \le \mathbbm{1}$. The transfer function then transforms the polarization of mode when mapping mode $m_i \to m'_i$ according to:
\begin{equation}
    \ket{\phi_\text{pol}'} = \frac{T_\text{pol}\ket{\phi_\text{pol}}}{||T_\text{pol}\ket{\phi_\text{pol}}||}.
\end{equation}

The scalar coefficients $T_i$ then correspond to the transmission amplitude, given by the norm of the filtered polarization component:
\begin{equation}
T_i = || T_\text{pol}\ket{\phi_\text{pol}}||.
\end{equation}
The remaining orthogonal component is transferred to the environmental mode $\hat{e}(m_i)$.

We note that this model extends to frequency or polarization dependent attenuation in lossy channels.

\subsection{Amplifying channels}

Amplifying devices constitute an important class of devices used in different communication aspects. Amplifying channels work similarly to the filtering channels, but in reverse. In this work we focus on general phase-insensitive optical amplifiers with constant spectral response, modeled by the map:
\begin{equation}
    \Phi_\text{amp}\left(\hat{a}(m'_i)\right) = \sqrt{G}\hat{a}(m_i) + \sqrt{G-1}\hat{e}^\dagger(m'_i), \quad G\geq1.
\end{equation}

In order to preserve the \gls{CCR}, it is required that any phase-insensitive amplifier introduces a coupling to the environmental modes. In the Bogoliubov map representation this corresponds to the case $V\neq0$, since, the expression:
\begin{equation}
    UU^\dagger - VV^\dagger = \mathbbm{1},
\end{equation}
forces $V\neq0$ for any gain $G>1$. The term $\sqrt{G-1}\hat{e}^\dagger$ represents the minimal quantum noise required by the quantum limit of amplification \cite{cavesQuantumLimitsNoise1982}.

To describe the state evolution by amplification, we switch to the Schrödinger picture. The above Heisenberg map arises from a unitary $U_\text{amp}$ acting jointly on the system and an environment.
\begin{equation}
    \rho'_{SE}
    = U_\text{amp}\left(\rho_S\otimes |0\rangle\!\langle 0|\right)
      U_\text{amp}^\dagger.
\end{equation}
Tracing out the environment yields the familiar phase-insensitive quantum-limited amplifier channel,
\begin{equation}
    \rho_S' = \mathrm{Tr}_E\big[\rho'_{SE}\big],
\end{equation}
a \gls{CPTP} map.
After establishing the induced state transformation, we return to the Heisenberg picture and continue to represent further device actions through operator maps on $\mathcal{A}_g$.

\subsection{Detection}
Lastly, we introduce the class of devices, which measure the quantum state. The detector is characterized by a \gls{POVM} $\{M_k\}$ acting on a chosen set of optical modes. The probability of each outcome $k$ is 
\begin{equation}
    p_k=\mathrm{Tr}(M_k\rho),
\end{equation}
and the corresponding post-measurement state is given by
\begin{equation}
    \rho_k' = \frac{(\sqrt{M_k}\otimes \mathbbm{1})\,\rho\,(\sqrt{M_k}\otimes 
    \mathbbm{1})^\dagger}{p_k},
\end{equation}
where the identity acts on the modes not measured. This ensures that a detection event on a single mode induces correct collapse of the global multimode state.

The two common detection processes are then implemented as follows:
\paragraph{Photon number detection} In the photon number detection the \gls{POVM} elements are set as $M_n=|n\rangle\!\langle n|$.
\paragraph{Threshold detection} For threshold detection the \gls{POVM} elements take form $M_\text{off}=\ket{0}\!\bra{0}$ and $M_\text{on}=\mathbbm{1}-M_\text{off}$.

Any detector inefficiency and spectral response can be naturally taken into account as a pre-pended filters prior to detection.

\section{Simulation Framework}

We formalize the mathematical constructions introduced in previous sections into a coherent simulation framework. The framework is designed to be a direct executable representation of the algebraic formalism.

The following paragraphs outline the core framework components.

\paragraph{Mode Representation} A mode label $m=(p,f)$ bundles together the spatial path $p$ with a single-photon state $f\in\mathcal{H}_1$, including the method for computing inner product. This encodes the mode orthogonality. 
All mode descriptors are stored as immutable objects. This ensures that label references always refer to fixed mathematical objects which in turn prevents hidden in-place value changes during transformations.

\paragraph{Operator Representation}
Ladder operators form the atomic building blocks for all expressions. Device actions are defined conceptually as rewriting rules that replace selected ladder operators according to the individual device map.
Ladder operators are also immutable, so they always represent creation or annihilation in exactly the same mode.
After rewriting procedure, operator products are normal-ordered using the \gls{CCR} which naturally produce contractions $\langle m_i, m_j\rangle$ whenever a creation operator passes an annihilation operator.

\paragraph{Operator Polynomial Representation}
States and observables are represented as finite sums of normally ordered monomials $M(I|J)$. 
Expectation values, traces, and partial traces reduce to evaluating vacuum contractions of these normally ordered expressions. The process follows from the rules established in Sec.~\ref{sec:operator_algebra}.
This representation naturally incorporates non-orthogonal modes.
\paragraph{Simulation Pipeline}
A simulation begins by constructing mode labels from the physical parameters of interest. These mode descriptors define the ladder operators, which can then be combined into operator polynomials representing the initial state. Device maps are applied at the algebraic level by replacing the affected operators according to their conceptual rewriting rules. Whenever an observable or trace is evaluated, the framework automatically performs normal ordering and carries out all required contractions. The workflow directly reflects the algebraic model developed in Sec.~\ref{sec:operator_algebra}.

The present prototype implements the full operator algebraic core for operator formation, normal ordering and contraction. It also implements a rewrite logic for passive linear optics. More general device classes follow the same interface and are left for future work. This is sufficient to present the proof-of-concept demonstration in this work.

\section{Hong-Ou-Mandel Simulation}\label{sec:hom_simulation}

To showcase the effectiveness of the proposed simulation framework, we analyze the famous \gls{HOM} experiment \cite{hongMeasurementSubpicosecondTime1987, raymerTemporalModesQuantum2020}. In the \gls{HOM} experiment, the indistinguishability of two photons entering a single (50/50) beam splitter determines the resulting interference. If the two photons are indistinguishable in all degrees of freedom, they always bunch and are both detected in the same output port of the beam splitter. If, however, there is any degree of distinguishability between the two photons, the interference is no longer perfect, leading to increased probability of photon detection at both output ports. This leads to the famous \gls{HOM} dip in the coincidence rate, where the coincidence corresponds to the detection event at both output ports.

In the proposed simulation framework, the two incoming photons are described as a quantum state:
\begin{equation}
    \ket{\psi} = \hat{a}^\dagger(m_1)\hat{a}^\dagger(m_2)\ket{0},
\end{equation}
where the two photons are initially distinguishable because they occupy different optical paths: $\langle p_1, p_2\rangle=0$, but they can still overlap in other degrees of freedom:
\begin{equation}
   \left|\bigl\langle \zeta_{T_1}, \phi_1 \mid \zeta_{T_2}, \phi_2 \bigr\rangle_{\mathcal{H}_1}\right|
\in [0,1]. 
\end{equation}

After the transformation due to the beam splitter, defined in Sec.~\ref{sec:linear_passive_devices},  the state then becomes:
\begin{equation}\label{eq:output_state}
   \ket{\psi'} = \frac{1}{2}\left(\hat{a}^\dagger\left(m'_{1, 1}\right)+\hat{a}^\dagger\left(m'_{1, 2}\right) \right)
   \left(\hat{a}^\dagger\left(m'_{2, 1}\right)-\hat{a}^\dagger\left(m'_{2, 2}\right) \right)\ket{0}.
\end{equation}

In this expression all four outcome possibilities are listed, the additional second index labels the optical output arm 1 or 2 of the beam splitter. The \gls{HOM} dip is hidden in the overlaps defined by the modes and the \gls{CCR} algebra.

To compute the coincidence of the two photons, we introduce the coincidence operator $\hat{C}$:
\begin{equation}
    \hat{C}=\hat{n}_{p'_1}\hat{n}_{p'_2},
\end{equation}
where individual number operators for each path are:
\begin{equation}
    \hat{n}_p=\sum_\alpha \hat{a}^\dagger(m_{\alpha, p})\hat{a}(m_{\alpha, p}),
\end{equation}
encompassing all possible modes in one particular path.
Then the coincidence expectation value can be computed with:
\begin{equation}
P_\text{coinc} = \langle \hat{C} \rangle_{\rho'} = \text{Tr}(\rho'\hat{C})=\langle \psi' | \hat{n}_{p_1}\hat{n}_{p_2}\mid \psi' \rangle = || \ket{\psi'_{1,1}}||^2.
\end{equation}

Computing the $P_\text{coinc}$ (carried out in appendix \ref{appendix:hom}) from this example yields exactly the \gls{HOM} dip formula:
\begin{equation}\label{eq:hom_dip}
    P_{\mathrm{coinc}} =\frac{1}{2}\left(1 - |\gamma|^2\right), \qquad
\gamma = \langle \zeta_{T_1}, \phi_1 \mid \zeta_{T_2}, \phi_2 \bigr\rangle_{\mathcal{H}_1}.
\end{equation}

\begin{figure}
    \centering
    \includegraphics[width=1\linewidth]{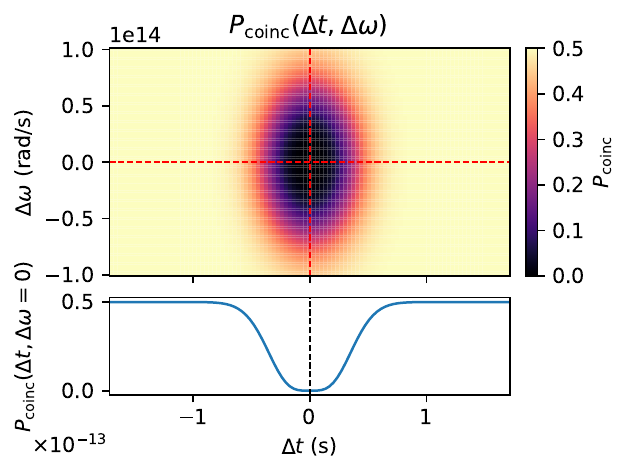}
    \caption{Simulated coincidence probability in relation to the time delay and spectral detuning $P(\Delta t, \Delta \omega)$. The \gls{HOM} dip occurs, as predicted, when the two incoming photons are indistinguishable in all modeled degrees of freedom.}
    \label{fig:simulation_hom}
\end{figure}

This analytical expression is reproduced directly by the simulation framework as displayed by the Fig.~\ref{fig:simulation_hom}. 

This simulation was performed using a lightweight symbolic simulation prototype. It illustrates how the operator algebra framework handles non-orthogonal temporal modes and polarization in a unified way.

\paragraph{Supplementary Material}
The source code and Jupyter notebook used to generate
Fig.~\ref{fig:simulation_hom} are available in the archived
release of the symbolic operator framework prototype~\cite{sekavcnik_symop_proto_2026}.

\section{Conclusions}
In this work, we presented a symbolic operator framework for quantum optical simulation, built on a Fock space construction and the \gls{CCR}. We formalized device maps that act directly on this \gls{CCR} algebra, and implemented a prototype simulator based on these rewrite rules. Using this formulation, the simulator successfully reproduces the \gls{HOM} interference dip and matches the analytical prediction.

Although the prototype is already equipped with the core machinery, future work will extend its capabilities to include abstraction as well as specific implementation of devices. We also plan to broaden the representation to fully support Gaussian and other continuous-mode states. In future work, computational complexity and possible optimizations will also be studied.

We are confident this approach provides a robust and extensible technique for modeling photonic systems with non-orthogonal modes, offering a compact operator formalism that remains faithful to the underlying optical physics.

\section*{Acknowledgments}
Authors are grateful to Florian Seitz for many valuable discussions within the TQSD research group.
\subsection*{Funding}
This work was supported in part by the DFG Emmy-Noether Program under Grant 1129/2-1, 
in part by the Federal Ministry of Research, Technology and Space under grant numbers 16KISQ039, 16KISR026, 16KIS1598K, 16KISQ093, 16KISQ077, and 16KIS2604 and in part by the Federal Ministry of Research, Technology and Space of Germany through the Programme of "Souverän. Digital. Vernetzt." Joint Project 6G-life, project identification number 16KISK002.
The generous support of the state of Bavaria via the 6GQT project is greatly appreciated. This research is part of the Munich Quantum Valley, which is supported by the Bavarian state government with funds from the Hightech Agenda Bayern Plus.

\appendix
\subsection{Contraction Examples}\label{appendix:contraction_examples}

To make the process of contraction and normal ordering clear we carry out some examples in this appendix.

\subsubsection{Inner Product}\label{appendix:inner_product}

Firstly, we illustrate the inner product between two states, 
\begin{equation}
    \ket{\psi_1} = \hat{a}^\dagger(m_1)\hat{a}^\dagger(m_2)\ket{0}, \quad 
    \ket{\psi_2} = \hat{a}^\dagger(n_1)\hat{a}^\dagger(n_2)\ket{0}.
\end{equation}

The inner product is then written as
\begin{equation}
    \langle\psi_1|\psi_2\rangle = \bra{0}\hat{a}(m_2)\hat{a}(m_1)\hat{a}^\dagger(n_1)\hat{a}^\dagger(n_2)\ket{0}.
\end{equation}

Using the \gls{CCR} rule the expression between $\bra{0}$ and $\ket{0}$ can be normal ordered, where the normal ordering produces additional terms with the inner product between modes:
\begin{equation}
\hat{a}(m_1)\hat{a}^\dagger(n_1)
= \hat{a}^\dagger(n_1)\hat{a}(m_1) + \ip{m_1}{n_1}\mathbbm{1}.
\end{equation}

Vacuum contractions survive only fully contracted terms \ie those in which all operators have been replaced by inner products. Any term with at least one remaining operator vanishes due to $\hat{a}\ket{0}=0$. In this example the resulting inner product then becomes:
\begin{align}
\langle \psi_1|\psi_2\rangle = \langle m_1|n_1\rangle \langle m_2|n_2\rangle
 +  \langle m_1|n_2\rangle \langle m_2|n_1\rangle.
\label{eq:two-photon-inner-product}
\end{align}

\subsubsection{Expectation value in Hong-Ou-Mandel experiment}\label{appendix:hom}
In the Sec.~\ref{sec:hom_simulation} we introduced a state after a 50/50 beam splitting action in Eq.~\eqref{eq:output_state}. Its component with one photon in each of the beam splitter arm is:
\begin{equation}
    |\psi'_{1,1}\rangle
    = \frac{1}{2}\left(\hat O_{12}\,|0\rangle - \hat O_{21}\,|0\rangle \right),
\end{equation}
with
\begin{align}
\begin{split}
    \hat O_{12} &= \hat a^\dagger(m_{1,1})\,\hat a^\dagger(m_{2,2}),\\
    \hat O_{21} &= \hat a^\dagger(m_{1,2})\,\hat a^\dagger(m_{2,1}).
\end{split}
\end{align}
Then computing its norm we get the coincidence probability:
\begin{equation}
    P_{\mathrm{coinc}}
    = \langle\psi'_{1,1}|\psi'_{1,1}\rangle
    = \frac{1}{4}\left(\bra{0}\big(\hat O_{12}^\dagger - \hat O_{21}^\dagger\big)
            \big(\hat O_{12} - \hat O_{21}\big)\ket{0}\right).
\end{equation}
\begin{align}
\begin{split}\label{eq:four_terms}
4P_{\text{coinc}}
      =& \bra{0}\hat O_{12}^\dagger \hat O_{12}\ket{0}
      - \bra{0}\hat O_{12}^\dagger \hat O_{21}\ket{0}\\
      &- \bra{0}\hat O_{21}^\dagger \hat O_{12}\ket{0}
      + \bra{0}\hat O_{21}^\dagger \hat O_{21}\ket{0}.
\end{split}
\end{align}
Then each term is normally ordered and contracted as in previous appendix example \ref{appendix:inner_product}. 
Denoting the overlap between internal modes in a given path by:
\begin{equation}
    \gamma = \ip{m_{1,1}}{m_{2,1}}
           = \ip{m_{1,2}}{m_{2,2}}.
\end{equation}
In the $\gamma$ expression, mode label $m_{a,p}$ denotes an internal single-photon mode attached to path $p$, so the overlap is taken only in the internal space. Equipped with $\gamma$ we then compute the values of individual terms:
\begin{align}
\bra{0}\hat O_{12}^\dagger \hat O_{12}\ket{0} &= 1,\\
\bra{0}\hat O_{21}^\dagger \hat O_{21}\ket{0} &= 1,\\
\bra{0}\hat O_{12}^\dagger \hat O_{21}\ket{0} &=|\gamma|^2\\
\bra{0}\hat O_{21}^\dagger \hat O_{12}\ket{0} &= |\gamma|^2.
\end{align}
Substituting back into \eqref{eq:four_terms}%
gives
\begin{equation}\label{eq:appendix_hom}
    P_{\mathrm{coinc}}
    = \frac{1}{4}\left(1 + 1 - 2|\gamma|^2
    \right)= \frac{1}{2}\left(1 - |\gamma|^2\right).
\end{equation}

\printbibliography

\end{document}